\documentstyle[aps,pra,amsfonts,epsfig]{revtex}

\begin{document} 


\title{Synchronization of rescaled adaptive coupling and its 
       application to shock capturing}

\author{G. W. Wei}
\address{Department of Computational Science, 
National University of Singapore\\
Singapore 117543, R. Singapore}

\date{\today} 
\maketitle

\begin{abstract}
 
This Letter proposes a rescaled  adaptive coupling scheme 
for the synchronization of spatially extended systems. 
Coupling and synchronization are analyzed from the point 
view of image filter construction. A length rescaling 
technique is introduced based on dimensional argument
of control process. Control sensors are adaptively 
selected according to local information. The coupling 
strength on each sensor is automatically adjusted 
according to the magnitude of local oscillations. 
We demonstrate that the present scheme can  efficiently
suppress and control spatiotemporal oscillations and thus,
provide a powerful approach for shock capturing.
Both the Navier-Stokes equation describing shear 
layer flows around a jet and Burgers' equation  
are solved to illustrate the idea.
\end{abstract} 

PACS numbers: 05.45.+b, 02.70.-c, 47.40.Nm


Synchronization phenomenon is of fundamental 
importance in telecommunication\cite{KocPar}, 
electronic circuits\cite{AVPS}, 
nonlinear optics\cite{FCR}, and
chemical and biological systems\cite{SchMar}.
The phenomenon has been studied extensively 
both by numerical and experimental means.
It is believed that an in-depth  study and understanding 
of synchronization will greatly benefit the 
advancement  of science and technology. 
Different types of 
synchronization, such as
identical\cite{FujYam,KocPar},
 generalized\cite{RST}, 
lag and phase  synchronization\cite{RPK},
were  proposed. Recently, synchronization and control of 
spatially extended systems have received great attention 
\cite{XHY,JunPar}. 
For a given system, the degree and rate of 
synchronization depends vitally on the coupling scheme
used. A variety of coupling schemes, such as unidirectional 
coupling, receptor-product coupling,  weak coupling,
strong coupling, global coupling, and local 
coupling, have been studied. However, adaptively 
coupled schemes and their effect on the rate and degree of 
synchronization have not been addressed yet.
Moreover, very little is reported on 
synchronization with respect to  the understanding of 
nonlinear hyperbolic conservation laws, shock capturing
and, in general, computational methodology. The latter
has had tremendous impact to a variety of disciplines
in  science and  engineering. 
In fact,  much of the 
present understanding on synchronization was achieved 
with the aid of numerical computations.

The main purpose of this Letter is to introduce  
rescaled adaptively
coupled synchronization schemes and to use them for 
shock capturing. A local gradient based coupling 
scheme is introduced for continuous systems. 
Nondecreasing functions and non-increasing functions 
are designed for oscillation reduction 
and image edge preservation, respectively. 
We demonstrate that appropriate coupling of 
two identical dynamical systems can result in
a surprisingly novel and efficient scheme for shock capturing. 
The validity and 
efficiency of this scheme is tested by using 
an inviscid Burgers' equation and the 
Navier-Stokes equation for incompressible flows.

For simplicity, we consider an identical synchronization,
where  two coupled systems are exactly of the same type 
and can be given by a partial differential equation of the from 
\begin{equation}\label{sync1}
u_t=F\left(u, u_x,u_{xx},
\cdots\right) +c(u,w),
\end{equation}                 
where $c(u,w)$ is a dissipative coupling term, which is 
proportional to the difference between the states of 
two systems, $(u-w)$.
Based on an experimental consideration, 
Junge and Partitz\cite{JunPar} 
proposed a sensor coupling scheme, which utilizes the 
difference between two  spatially averaged local  signals 
$(\bar{u}-\bar{w})$.  Here,
$\bar{w}(x,t)={1\over l}\int_{x-l/2}^{x+l/2}w(y,t)dy$
is the local average of $w$ over a length $l$ at the position 
of a sensor. 
It is noted that a dimensional 
scaling effect has to be accounted  when the solution 
of a partial differential 
equation is related to an experimental measurement. 
In fact, in control experiments, two coupled systems 
might have very different spatial extensions 
(the control 
part is usually much smaller in size, e.g., 
a human brain vs human body). Mathematically, the 
dimensionless partial differential equation usually 
has one or more parameters which contain the length
scale. We argue that, when two dynamical systems 
characterized by different values of a parameter 
are coupled together, one of the two dynamical systems 
should be appropriately rescaled. Therefore,
we propose a rescaled coupling scheme
\begin{equation}\label{sync2}
c(u,w) \propto (u-\bar{w}),
\end{equation}
where $\bar{w}$ is a local average of $w$.
It is important to understand that the coupling 
between two  systems given by 
Eq. (\ref{sync2}) is generally designed as a 
dissipative coupling. However, an interesting observation 
can be made at the limit of complete identical
synchronization (i.e. the strong convergence of the two systems
$\| u(t)-w(t)\| \rightarrow 0$ for $t\rightarrow\infty$).
From the point of view of image processing,
the local average $\bar{w}$ is equivalent to the treatment of 
${w}$  by a low-pass filter. Moreover, at the limit of 
complete identical synchronization,  ($u-\bar{w}$) 
is equivalent to the treatment of 
$u$ by a high-pass filter\cite{GonWoo}. 
There is a similar effect on the second 
system under the same condition. As such, 
the rescaled sensor coupling expression given by Eq. (\ref{sync2})
can be used for image processing, pattern recognition 
and shock capturing. In these applications, spatially 
selected treatment is of practical importance. To achieve 
spatial selectivity, we introduce  following 
rescaled and adaptively distributed local sensors
\begin{equation}\label{sync3}
c(u,w) \propto \varepsilon(\mid u_x\mid) (u-\bar{w}),
\end{equation}
where the coupling strength $\varepsilon$ 
is a function of the gradient measurement 
$\mid u_x\mid$.
For the purpose of edge-detected pattern recognition, 
 we  choose $\varepsilon(\mid u_x\mid)$ as 
a non-increasing function, e.g.  
\begin{equation}\label{sync4}
\varepsilon(\mid u_x\mid)=\epsilon\exp\left[
-{\mid u_x\mid^2\over2\sigma^2}\right],
\end{equation}       
where $\epsilon$  and $\sigma$ are constants.
For the purpose of noise reduction and oscillation 
suppression, we  choose $\varepsilon(\mid u_x\mid)$ as 
a nondecreasing function, e.g.  
\begin{equation}\label{sync5}
\varepsilon(\mid u_x\mid)=
\epsilon\mid u_x\mid^{1\over4}, 
\end{equation}       
where $\epsilon$ is a constant. Obviously many other 
nondecreasing functions can also be used.
In the rest of the Letter we restrict ourselves to the 
application of  the present synchronization scheme to
shock capturing.

The solution of the inviscid Burgers' equation and the 
incompressible Navier-Stokes equations for very high 
Reynolds numbers is often difficult to attain
due to the possible existence of shock front. 
Shock wave is a common phenomenon in nature, such as 
in aerodynamics and hydrodynamics, and is usually 
described by hyperbolic conservation laws and by 
inviscid  hydrodynamic equations.
The construction of numerical schemes that are 
capable of efficient shock capturing is a challenging task.

To illustrate our synchronization approach for
oscillation reduction,  we first consider 
Burgers' model of turbulence
\begin{equation}\label{Burgers}
u_t+u u_x
={1\over {\rm Re}}u_{xx},
\end{equation}                 
where $u(x,t)$ is the dependent variable 
resembling the flow velocity and Re is the 
Reynolds number characterizing the size of the viscosity.
Burgers' equation is an important model 
for the understanding of physical flows.
The competition between the nonlinear advection and the
viscous diffusion is 
controlled by the value of Re in Burgers' equation, and
thus determines the behavior of the solution. 
 We consider Eq. (\ref{Burgers}) using the 
following initial and boundary conditions
\begin{eqnarray}\label{bur2}
u(x,0)=\sin(\pi x),~~
~~u(0,t)=u(1,t)=0.
\end{eqnarray}
The forth order Runge-Kutta scheme is used for the temporal 
discretization with a time increment $\Delta t =0.002$.
A discrete singular convolution (DSC) 
algorithm\cite{weijcp99,weijpa20} 
is utilized for spatial 
discretization with a total of 101 grid points
in the computational domain. The use of DSC algorithm 
for solving differential equations has been extensively 
tested and further validation is given in TABLE I.

Solving Burgers' equation at a high Reynolds number is 
a challenging task. At a Reynolds number of $10^3$,
the numerical solution quickly 
develops into a sharp shock front near $x=1$.
Severe oscillations occur near the 
shock front as shown in FIG. 1a. 
It should be pointed out that, almost all high order 
numerical schemes exhibit similar oscillations.
To eliminate oscillations, we employ the
rescaled adaptive coupling, Eq.(\ref{sync5}), in solving 
Burgers' equation (\ref{Burgers}).
Here $\bar{w}$ is computed by a local three-term average.
Two systems, which are characterized by 
two Reynolds numbers (Re$_1=1000$ and Re$_2=100$),
are coupled with a coupling constant of $\epsilon=-80$.  
It can be  seen from FIG. 1b that
all spurious oscillations could be eliminated. 
However, the synchronized 
solution is neither the true solution of Re$=100$
nor that of Re$=1000$. Hence,  it is desirable to have an
oscillation-free solution at a given 
high Reynolds number. To this end, we design 
an {\it autosynchronization} approach 
by choosing two exactly identical systems, i.e., 
setting Re$_1$ =Re$_2=\infty$.
As two exactly identical systems are still coupled, 
oscillations are suppressed to a certain degree, 
depending on the coupling constant. 
For a relatively small coupling constant of $\epsilon=-40$, 
the solution is  oscillatory at early times and
become  essentially non-oscillatory at a later time (see FIG. 1c).
By increasing the coupling constant to $\epsilon=-90$,
we have successfully eliminated all spurious 
oscillations as shown in FIG. 1d.

To validate the present approach further,
we consider the two dimensional Navier-Stokes equation
\begin{eqnarray}\label{NS1}
{\bf U}_t+ {\bf U\cdot \nabla U}=-\nabla p+{1\over {\rm Re}}\nabla^2{\bf U}
+\varepsilon(\mid {\bf\nabla U}\mid)({\bf U}-\bar{\bf W})
\end{eqnarray}
with the incompressible condition, ${\bf\nabla \cdot U}=0$.
Here ${\bf U}=(u,v)^T$ is the 
velocity vector, ${\bf W}$ is the velocity vector
of the second system,  $p$ is the pressure, 
Reynolds number of Re$ =\infty$ defines the Euler equation. 
The domain of interest is 
a square $[0,2\pi]\times [0,2\pi]$ with
periodic boundary conditions. 
Depending on the initial values, this system can be 
very challenging to solve. 
For smooth initial values,
the solution scheme and the validity of the DSC 
algorithm were tested in Ref. \cite{weijpa20}. 
With appropriate initial values, the Euler 
equation (Re$ =\infty$) can be used to describe 
the flow field of vertically perturbed
horizontal shear layers around a jet.

We now test our synchronization approach 
for the Euler equation 
with sharply varying initial values.
This example is chosen to 
illustrate the ability of the present approach 
for providing very fine resolution even on a 
relatively coarse grid. 
The initial values are that of a jet in a doubly 
periodic geometry
\begin{eqnarray}\label{NS7}
&&u(x,y,0)=\left\{\begin{array}{ll}\mbox{$
\tanh\left({2y-\pi\over 2\rho}\right),$} & \mbox{if $y\leq\pi$}\\
\mbox{$\tanh\left({3\pi-2y\over2\rho}\right),$} & \mbox{if $y >\pi$}
\end{array}
\right\} \nonumber\\
&&v(x,y,0)=\delta\sin(x),
\end{eqnarray}
where $\delta=0.05$ is used for the convenience of 
comparison with the previous study\cite{BelCol,EShu}.
This initial value describes the flow field consisting of 
horizontal shear layers of finite thickness, perturbed by a small 
amplitude vertical velocity, making up the boundaries of the jet. 
However, this problem is not analytically solvable.
Pioneering work  was done  by 
Bell et al\cite{BelCol} in this field, in which 
they utilized a second-order Godunov 
scheme in association with 
a projection approach for divergence-free velocity field
with a general boundary condition. 
A state of the art high-order  essentially non-oscillatory 
(ENO) scheme was constructed by E and Shu\cite{EShu} to 
resolve the fine vorticity structure  
of the double shear layers with periodic boundary conditions.

We consider the parameter  $\rho=\pi/15$, 
a case studied by Bell et al\cite{BelCol} using 
a projection method with three sets of grids
(128$^2$, 256$^2$ and  512$^2$).
E and Shu\cite{EShu} computed this case by 
using both  spectral collocation code 
with 512$^2$ points and their high order ENO scheme
with 64$^2$ and  128$^2$
points. The spectral collocation code
produced an oscillatory 
solution at $t=10$ (see FIG. 1 of Ref. \cite{EShu}), 
while the high order ENO scheme
produced a defect at $t=6$ as the channels 
connecting the vorticity centers are slightly
distorted (see FIG. 2 of Ref. \cite{EShu}).
In the present simulation, we
choose a $64^2$ grid for the 
computational domain with a time increment of 0.002.
The synchronization prescription given 
in Eq. (\ref{sync5}) is used with 
a coupling constant of $\epsilon=-80$. 
The results at different times ($t$=4,6,8, and 10)
are plotted in FIG 2. 
It is seen that our solution is smooth (some non-smooth
features in the contour plot are attributed to the coarse 
grid employed in the study) and stable for this case. 
In particular, no distortion is found in vorticity 
contours at t=6.
For early times, present results compare extremely 
well with those of the spectral collocation code computed 
with 512$^2$ points. There are no spurious
numerical oscillations during the entire process.

In conclusion, we propose the approach of  synchronization 
as a robust, reliable and practical algorithm for 
shock wave computations. 
A length rescaling technique is introduced 
based on the dimensional
argument of control process. To achieve computational 
efficiency and reliability, coupling sensors in 
spatially extended systems
are adaptively selected according to 
local information.
The coupling strength at each sensor is adaptively varied 
according to the magnitude of the local gradient of the system. 
The resulting coupled systems 
are analyzed from point of view of image filters.
The proposed algorithm is validated by using Burgers' equation
and the incompressible Navier-Stokes equation.
A high accuracy  discrete singular convolution 
algorithm \cite{weijcp99,weijpa20}
is utilized for the numerical simulation.

For  Burgers' equation, we have tested the computational 
accuracy and reliability at a moderately high Reynolds number.
At high Reynolds numbers, Burgers' equation is difficult
to solve. We first test the synchronization approach 
in two systems for different Reynolds numbers. One of the two 
systems develops  spurious oscillations which could be  
eliminated by coupling to a low Reynolds number  system.
The oscillations are completely suppressed at synchronization.
To make the algorithm practical for shock capturing at any 
given Reynolds number,
we couple two truly identical systems (i.e., the same 
dynamical system, the same parameters). It was found that  the
oscillations could be completely eliminated
above a suitable minimum coupling strength.

To further validate the present synchronization approach 
for shock capturing in  practical large scale 
computations, we consider the simulation of 
doubly periodic shear layer flows. For this purpose,
the Navier-Stokes equation with appropriate initial 
and boundary conditions is employed.
Numerical simulation of this system is an acid test for
ordinary methods, including  spectral methods. 
Our synchronization scheme performs extremely well.
With a relatively small grid mesh, the present results
are better than those of a state of the art shock capturing 
scheme, the high order essential non-oscillatory (ENO) 
scheme\cite{EShu}, obtained over a much larger grid mesh.
This indicates that the proposed approach has a great potential 
for being used as an efficient and reliable algorithm 
for the practical simulation of fluid flows
and computational physics in general.
Progress has been made on the application of 
the present synchronization approach to 
image processing.

{This work was supported in part by the 
National University of Singapore.}




\begin{table}
\caption{ $L_1$  and $L_\infty$  
errors of the DSC solutions for 
           Burgers' equation}
\begin{center}
\begin{tabular}{c|c|c|c|c|c|c|c} 
$t$  &  0.4  & 1.2 & 2.0 &  3.0 & 10  & 20  &   40  \\ \hline
$L_{1}$ &  2.1(-4) & 4.2(-5) & 9.4(-7) & 3.8(-8) & 1.5(-11)&
 3.2(-12)& 4.1(-13)\\ 
$L_{\infty}$ &  2.8(-3)&  6.8(-4) &  1.2(-5)&  4.0(-7)
&  4.0(-11))&  5.3(-12)&  6.4(-13)
\end{tabular}
\end{center}               
\end{table}

\vskip 30pt

\centerline{\bf Figure Captions}

{\bf FIG. 1.} Synchronization profiles of Burgers' equation at
$t=$0.2 (i), 0.4 (ii), 0.6 (iii), 1.2 (iv) and 2.0 (v).
(a) $\epsilon=0$, solid line: Re$_1$=1000, dots: Re$_2$=100; 
(b) $\epsilon=-80$, solid line: Re$_1$=1000, dots: Re$_2$=100; 
(c) $\epsilon=-40$, Re$_1$=Re$_2$=$\infty $; 
(d) $\epsilon=-90$, Re$_1$=Re$_2$=$\infty $.

\vskip 20pt

{\bf FIG. 2.} 
The vorticity contours of the synchronization solution of 
the 2D Euler equation with 
64$^2$ points.
Up left:   $t=4$;
up right:  $t=6$;
low left:  $t=8$;
low right: $t=10$.

\newpage

\begin{center}
\resizebox{14cm}{!}{\includegraphics{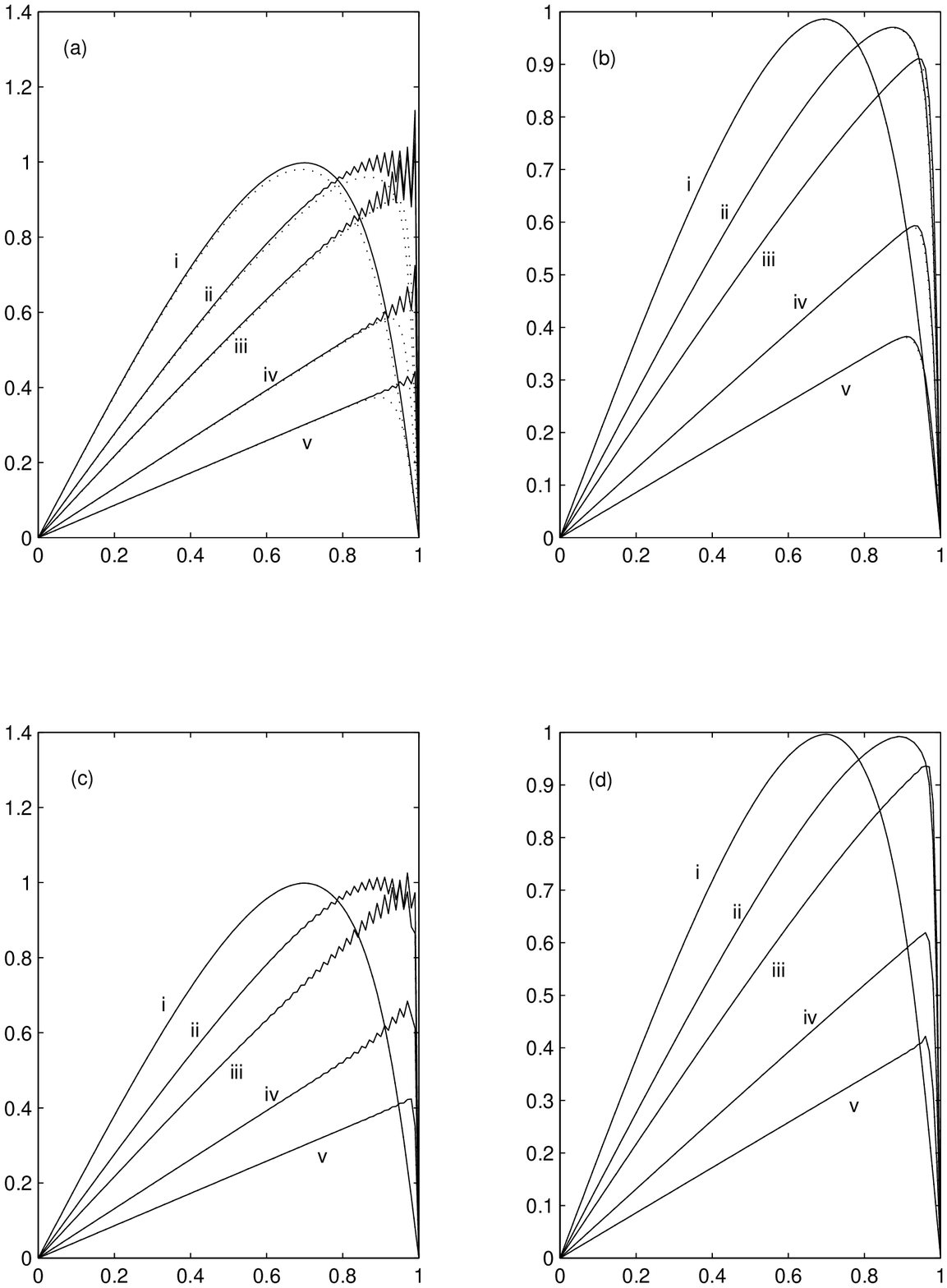}}
\vskip 1cm
FIG. 1
\end{center}

\newpage

\begin{center}
\resizebox{14cm}{!}{\includegraphics{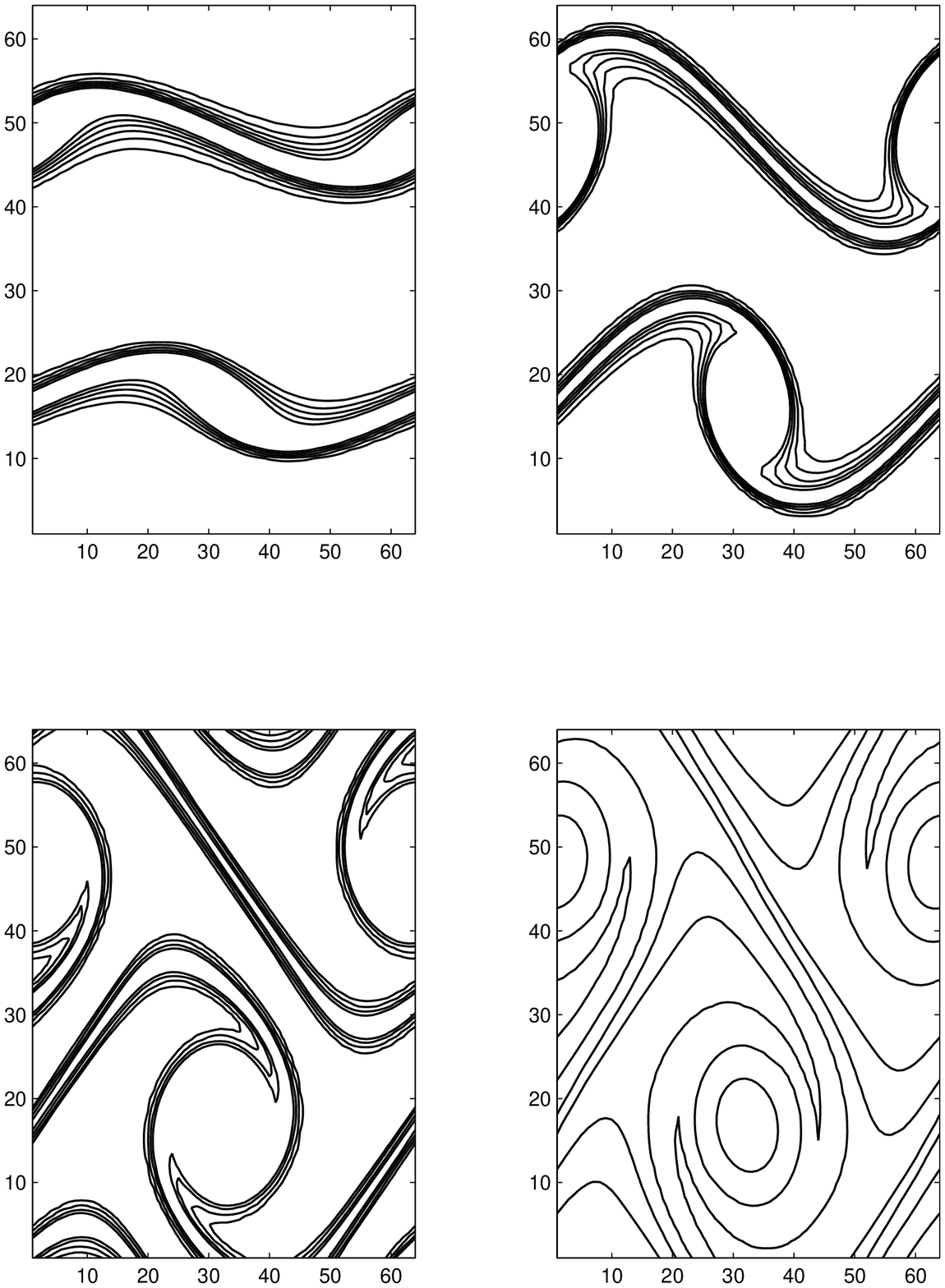}}
\vskip 1cm
FIG. 2
\end{center}


\begin{thebibliography}{99}

\bibitem{KocPar} L. Kocarev and U. Parlitz, Phys. Rev. Lett. {\bf 74}, 
           5028 (1995).

\bibitem{AVPS}V. S. Anischenko, T. E. Vadivasova, D. E. Postnov, 
       and M. A. Safonova, Int. J. Bifurcation Chaos Appl. Sci. Eng.
       {\bf 2}, 633 (1992); J. F. Heagy, T. L. Caroll, and L. M. Pecora,
        Phys. Rev. A {\bf 50}, 1874 (1994). 

\bibitem{FCR}L. Fabiny, P. Colet, and R. Roy, Phys. Rev. A {\bf 47},
        4287 (1993); R. Roy and K. S. Thornburg, Jr., Phys. Rev. Lett.
        {\bf 72}, 2009 (1994).


\bibitem{SchMar}I. Schreiber and M. Marek, Physica (Amsterdam) 
        5{\bf D}, 258 (1982); S. K. Han, C. Kurrer, and 
       Y. Kuramoto, Phys. Rev. Lett. {\bf 75}, 3190 (1995).


\bibitem{FujYam}H. Fujisaka and T. Yamada, Prog. Theor. Phys. 
       {\bf 69}, 32 (1983).

\bibitem{RST}N. Rulkov, M. Sushchik, L. Tsimering, and H. Abarbanel, 
             Phys. Rev. E {\bf 51}, 980 (1995);
         L. Kocarev and U. Parlitz, Phys. Rev. Lett. {\bf 76}, 
           1816 (1996).

\bibitem{RPK}M. G. Rosenblum, A. Pikovsky, and J. Kurths,
              Phys. Rev. Lett. {\bf 76}, 
           1816 (1996); U. Parlitz, L. Junge, W. Lauterborn, 
          and L. Kocarev, Phys. Rev. E {\bf 54}, 2115 (1996).


\bibitem{XHY}J. H.  Xiao, G. Hu, J. Z. Yang, and J. H. Gao,
        Phys. Rev. Lett. {\bf 81}, 5552 (1998); 
        S. Boccaletti, J. Bragard,  F. T. Arecchi, and  H. Mancini,
        Phys. Rev. Lett.
        {\bf 83}, 536 (1999); B. Blasius, A. Huppert, and L. Stone,
          Nature {\bf 399}, 354 (1999). 


\bibitem{JunPar} L. Junge and U. Parlitz,  Phys. Rev. E 
          {\bf 61}, 3736 (2000).

\bibitem{GonWoo}R. G. Ganzalez and R. E. Woods, 
        {\it Digital Image Processing}, p196, 
        (Addison-Wesley Company, Inc. New York, 1993).  


\bibitem{weijcp99} G. W. Wei,
        J. Chem. Phys. {\bf 110}, 8930 (1999);
       G. W. Wei, 
       Physica D {\bf 137}, 247 (2000);        
        G. W. Wei, 
         J. Phys. B {\bf 33}, 343 (2000).  


\bibitem{weijpa20} G. W. Wei, 
        J. Phys. A, {\bf 33}, 4935 (2000); G. W. Wei, 
         Computer Meth. Appl. Mech. Engng. in press.



\bibitem{BelCol}J. B. Bell, P. Cilella and H. M. Glaz,
            J. Comput. Phys. {\bf 85}, 257 (1989).

\bibitem{EShu} W. E and C.-W. Shu, 
       J. Comput. Phys., 
      {\bf 110}, 39 (1994).








\end{thebibliography}
\end{document}